%% file: main-arxiv.tex
\documentclass[11pt]{article}

\usepackage[utf8]{inputenc}
\usepackage[T1]{fontenc}
\usepackage{lmodern}
\usepackage{microtype}
\usepackage[margin=1in]{geometry}
\usepackage{graphicx}
\usepackage{subcaption}
\usepackage{booktabs}
\usepackage{hyperref}

\usepackage[section]{placeins}

\usepackage[backend=biber,style=numeric,sorting=none]{biblatex}

\title{Lightfall: An API-first, LLM-addressable control platform for synchrotron beamlines}

\author{%
  Ronald J.~Pandolfi$^{1}$, Damian Guenzing$^{2}$, Marcus M.~Noack$^{1}$, \\
  Sophie A.~Morley$^{2}$, Damon English$^{2}$ \\[6pt]
  \normalsize $^{1}$Applied Mathematics and Computational Research Division, \\
  \normalsize Lawrence Berkeley National Laboratory \\
  \normalsize $^{2}$Advanced Light Source, Lawrence Berkeley National Laboratory%
}

\date{June 2026}

\begin{document}
\maketitle

\begin{abstract}
\input{content/00-abstract}
\end{abstract}

\input{content/01-introduction}
\input{content/02-architecture}
\input{content/04-llm-as-designer}
\input{content/03-llm-as-user}
\input{content/05-supporting}
\input{content/06-deployment}
\input{content/07-discussion}
\input{content/08-conclusions}
\input{content/09-acknowledgments}

\printbibliography

\end{document}

%% file: content/00-abstract.tex
Synchrotron beamlines differ in hardware, technique, and workflow,
making customized control interfaces necessary; bespoke per-beamline
graphical user interfaces (GUIs) do not scale well, one-size-fits-all
facility software forces
compromises that leave most of the interface unused, and even
recent component-library approaches keep per-scientist tweaks
on a developer's queue.
We present Lightfall,
a control platform designed for facility-wide use, whose API-first
architecture
exposes every panel, device, and scan plan through a single uniform
addressable interface.
An embedded language-model agent drives experiments through that
interface, from a single move-and-read to a Gaussian-process-driven
autonomous scan, while beamline staff extend the interface during
operation via \emph{skills}: plugin modules the agent invokes to
compose and modify panels in the running application.
The result is a closed development loop: a beamline scientist
authors a panel change in natural language, the agent emits and
applies it, and the commit lands in the beamline's plugin
repository as a side effect. The per-iteration cost of a
scientist-driven change is then fixed in the scientist's own time
rather than in developer hours the facility must supply.
Lightfall is in testing at the COSMIC-Scattering beamline at the
Advanced Light Source.

%% file: content/01-introduction.tex
\section{Introduction}\label{sec:introduction}

Every synchrotron beamline is, in a practical sense, its own instrument.
Hardware configurations differ across beamlines (detector geometry, motion
stages, photon-delivery optics), and so do the experimental techniques and
the communities of users those techniques serve.
A generic facility graphical user interface (GUI) designed to cover the broadest common case inevitably
leaves most of its interface unused at any given beamline, while forcing the
beamline scientist to work around assumptions that do not match local
practice.
The alternative, building a bespoke interface for each beamline, produces
software that fits the immediate need but rarely survives the departure of
its author: the code accumulates undocumented workarounds, dependencies
drift, and the next staff member inherits a system they did not design and
cannot easily maintain.
This tension between facility-wide consistency and beamline-specific fit is
not new to the synchrotron community; it has shaped control and analysis
software for decades.
Xi-CAM \cite{pandolfi2018xicam} confronted an analogous tension across data
acquisition, orchestration, and analysis, and showed that a plugin
architecture can absorb some of it. The harder constraint Xi-CAM exposed was
development momentum rather than architectural rigidity: a single-developer
team cannot practically sustain the long tail of domain-specific and
bespoke user interfaces that a multi-technique facility accumulates, even
when a plugin model lowers the per-plugin cost.
Bundling many domain-specific concerns (much of it data analysis) into a
single process also made dependency management a recurring drag on a
small team. The plugin contract reduced the cost of adding a feature; it
did not change who had to author it.
A broader account of the mathematics, algorithms, and software infrastructure
required by modern experimental facilities is given by Chang
et~al.~\cite{chang2020building}; Xi-CAM and Lightfall are two contributions
to that broader program.

Prior work in beamline control software occupies the same spectrum without
closing the gap.
LabVIEW-based tooling\footnote{The Beamline Control System (BCS),
maintained by the ALS controls group; facility-internal software with
no standalone publication, described here for context.} underpins
acquisition at most beamlines at the
Advanced Light Source (ALS), where this work was developed; it is less
common at other facilities. At the ALS the LabVIEW layer is a unified,
facility-wide interface in its own right, shared across beamlines and
deliberately scoped to acquisition in order to keep the per-beamline cost
of joining the platform manageable. In that respect it is comparable to
Diamond's GDA \cite{opengda} and CSS-Phoebus \cite{css-phoebus}: a
facility-consistent environment that can be deployed uniformly across an
accelerator complex, but whose customization model requires forking display
files or writing scripts that sit outside the main control flow, and whose
scope is bounded in order to keep that model tractable.
Technique-specific tools such as Blu-Ice \cite{mcphillips2002bluice} and
MxCuBE \cite{gabadinho2010mxcube} achieve genuine usability for their target
domain (macromolecular crystallography in both cases), but their design
assumptions do not generalize to other techniques.
A more recent response to the same tension, exemplified by
Finch~\cite{deleon2025finch}, replaces the per-beamline GUI with a
shared component library that beamline developers import to assemble
per-beamline web applications. The per-beamline cost shifts from
``forking a GUI'' to ``composing UI atoms,'' but the customization
loop retains its shape: each beamline maintains its own application
repository, per-scientist tweaks are authored by developers
rather than by the scientist, and per-beamline application code
becomes the long-term maintenance surface. Total facility effort
remains tied to developer-hours per scientist-driven iteration:
the same scaling profile a fully forked GUI imposes, with a smaller
leading constant.
Each of these approaches addresses part of the problem.
Prevalent beamline software does not let a beamline scientist close
their own iteration loop
(author a panel modification, commit it to the beamline's plugin
repository, and have it live in the running interface) without
routing through a developer. That constraint is what determines
whether a customization model scales as the facility grows: whether
the marginal cost of a scientist-driven change is fixed in the
scientist's own time, or grows linearly in developer effort the
facility must supply.

The emergence of large language models (LLMs) with structured tool use
\cite{yao2023react,anthropic-tool-use} changes what a control interface can
be.
By \emph{agent} we mean, throughout this paper, a program in which an LLM
does more than generate text: the model is given a set of callable
\emph{tools} (structured API operations with typed parameters), chooses
which to invoke, observes the results, and iterates until the user's
request is satisfied.
When every panel, device, and scan plan is exposed through a uniform,
machine-readable application programming interface (API), the interface
becomes addressable by such an agent: it can query the current state of
the beamline, invoke a scan, or compose
a new display element by making the same API calls a human operator would.
Customization then becomes a matter of composing API calls in natural
language rather than writing GUI code.
Recent prototypes have begun to demonstrate this approach at scientific
user facilities: VISION~\cite{mathur2025vision} shows that a modular,
LLM-driven natural-language assistant can mediate scientist--instrument
interaction at a synchrotron, focused on instrument control as the primary
agent role.
Exposing the control interface to agents in this way reframes the scaling
problem: the bottleneck moves from
developer-hours-per-beamline to the design quality of the underlying API.
A well-designed API amortizes across every beamline that adopts the platform;
a well-designed agent amortizes across every task that can be expressed
through it.
The two interact: each improvement to the API makes the agent more capable,
and each new agent capability reduces the demand for hand-written tooling.

Lightfall demonstrates that a single API-first design can make a beamline control
system addressable by an LLM in two complementary roles (as a user of the
interface and as a developer of it), yielding a sustainable alternative to
both bespoke per-beamline GUIs and one-size-fits-all facility software.
The remainder of this paper is organized as follows.
Section~\ref{sec:architecture} describes the API-first architecture that
makes this addressability possible.
Section~\ref{sec:llm-designer} shows beamline staff extending the interface
through runtime skills the agent invokes; this is the paper's headline
result.
Section~\ref{sec:llm-user} shows the embedded agent driving experiments
through the interface in its role as a user.
Section~\ref{sec:supporting} covers supporting capabilities that the same
architecture enables (FAIR-compliant data access and a unified user
experience) and reports the external tools that already drive Lightfall
through its programmatic interfaces.
Section~\ref{sec:deployment} reports deployment at the COSMIC-Scattering
beamline at the Advanced Light Source.
Section~\ref{sec:discussion} discusses limitations and directions for future
work, and Section~\ref{sec:conclusion} concludes.

%% file: content/02-architecture.tex
\section{Architecture}\label{sec:architecture}

Lightfall is organized around three design principles that together make the
system addressable by both human operators and software agents.
First, \emph{API-first}: every user-facing surface has a programmatic
representation that can be queried and invoked independently of the GUI
that renders it.
Second, \emph{plugin-extensible}: the system is composed of typed plugins
(panels, settings, engines, plans, and skills) that register against a
shared runtime; no beamline-specific behavior requires modifying core
code.
Third, \emph{progressive disclosure}: routine controls are visible to all
users, while privileged or expert operations sit behind authentication
boundaries enforced at the API layer rather than hidden in the GUI.
These principles are not independent; API-first design is what makes
plugin extensibility tractable, and authentication boundaries are only
meaningful if every entry point goes through the API.

\begin{figure}[tbp]
  \centering
  \includegraphics[width=0.95\linewidth]{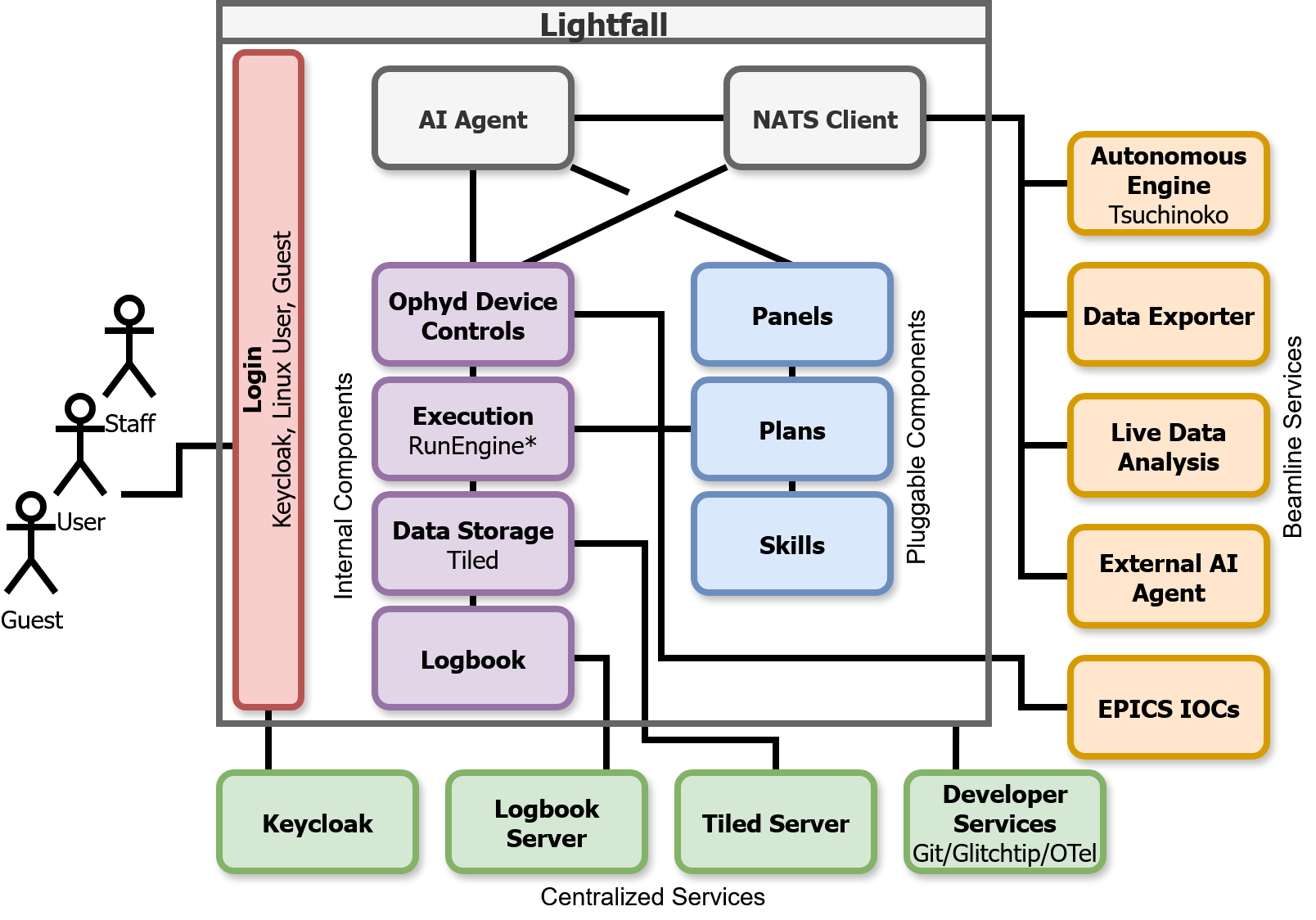}
  \caption{Lightfall architecture. Staff, scientist, and guest users
  authenticate through a Keycloak-backed login and interact with the Lightfall
  runtime. Internal components (purple) cover Ophyd device controls, the
  Bluesky RunEngine execution layer, Tiled-backed data storage, and the
  logbook; the embedded AI agent and NATS client address these components
  through the same interface the GUI uses. Pluggable components (blue),
  covering panels, plans, and skills, register against that surface.
  Centralized services (green) provide Keycloak, the Lightfall logbook server,
  the Tiled server, and developer tooling (git, Glitchtip, OpenTelemetry).
  Beamline services (orange), including autonomous engines such as Tsuchinoko,
  data exporters, live analysis, external AI agents, and EPICS IOCs,
  reach the system through the NATS fabric.}
  \label{fig:arch}
\end{figure}

\begin{figure}[tbp]
  \centering
  \includegraphics[width=0.95\linewidth]{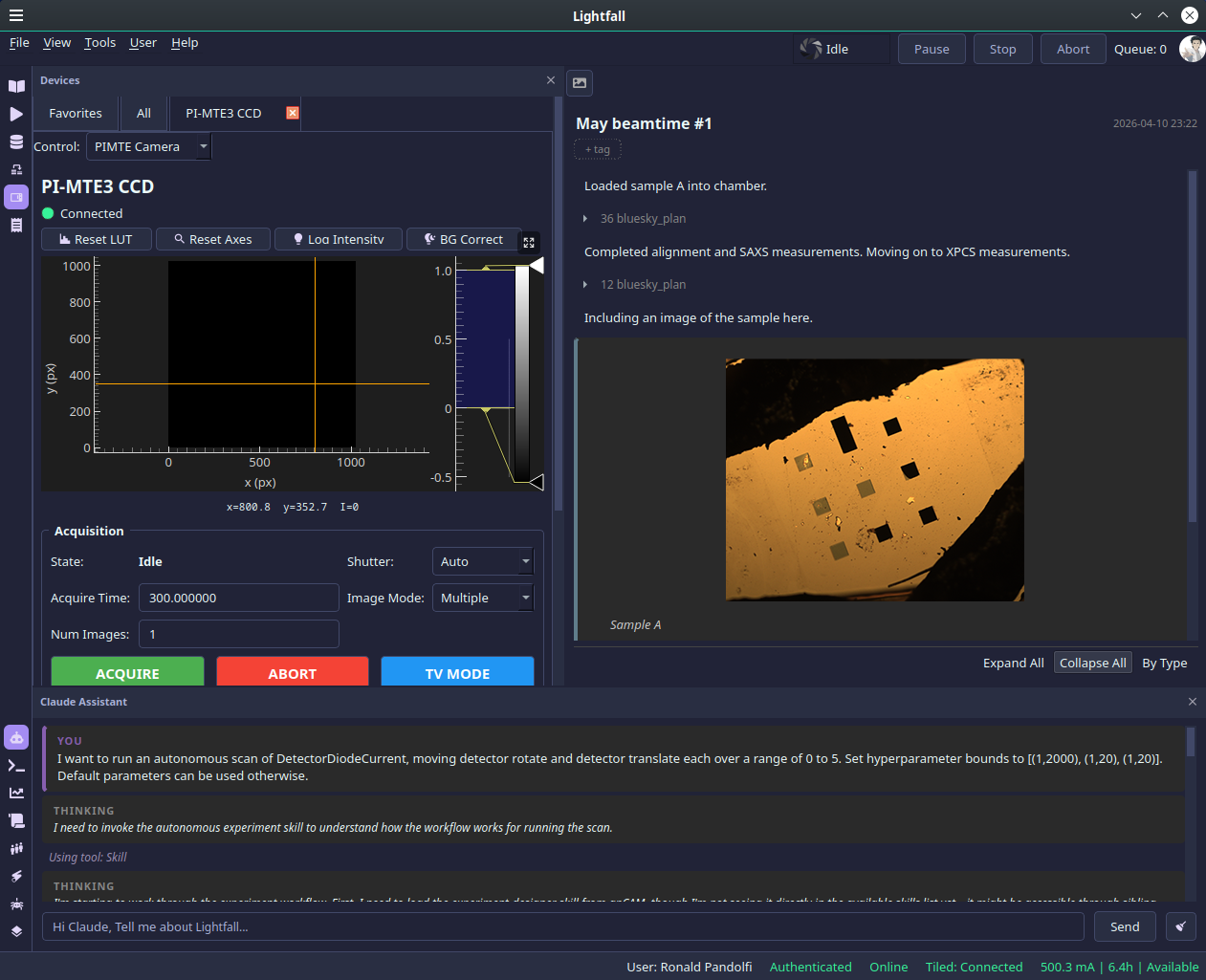}
  \caption{Lightfall in operation at the COSMIC-Scattering beamline
  (deployment is described in \S\ref{sec:deployment}).
  Center: a live PI-MTE3 CCD device panel with acquisition controls.
  Right: the beamline logbook capturing a beamtime in progress, with
  free-text notes from the scientist interleaved with collapsed
  Bluesky-plan invocation groups and embedded analysis snapshots,
  building a single timestamped record of the session. Bottom: the
  embedded agent receiving a control-mode request to launch an
  autonomous scan, with the scientist's natural-language prompt and
  the agent's reasoning trace inline. Upper right: RunEngine controls
  (pause, stop, abort), the plan-queue indicator, and the avatar of the
  Keycloak-authenticated user session through which every action in the
  window is authorized. Status bar (lower edge):
  authenticated principal, Tiled connection state, and live storage-ring
  status (current, beam lifetime, availability). One application
  surface threads through the control-mode and design-mode agents, the
  device interface, the data catalog, and the logbook, with each
  contribution attributable to its origin in a single audit trail.}
  \label{fig:cosmic}
\end{figure}

Figure~\ref{fig:arch} shows these layers together: scientists and staff
enter the uniform API through the GUI or agent; panels, devices, and plans
are peers on the middle layer; plugins register here, not against the
renderer, ensuring every capability is agent-addressable at installation.
Figure~\ref{fig:cosmic} shows the same layers realized in the running
application as deployed at the COSMIC-Scattering beamline; the deployment
itself is described in \S\ref{sec:deployment}.

\subsection{Core stack}\label{subsec:core-stack}

Lightfall builds on the Bluesky ecosystem \cite{allan2019bluesky}, which has
seen recent increased adoption for experiment orchestration at synchrotron
facilities internationally (NSLS-II, APS, ALS, Diamond, and BESSY~II among
them); that uptake means hardware and scan procedures at participating
facilities already speak a common language, enabling cross-facility
autonomous experiments of the kind recently demonstrated at the ALS and
PETRA~III \cite{koepp2026unified}.
Experiment sequencing is handled by Bluesky's RunEngine; device
abstraction is provided by Ophyd \cite{ophyd}, which presents every
hardware component (motors, detectors, shutters) as a uniform Python
object with a consistent command surface regardless of the underlying
hardware protocol.
Data is cataloged by Tiled \cite{tiled}, a service with a standard web
(REST) interface that
makes run data queryable and streamable. EPICS \cite{epics} provides the
channel-access backbone; the GUI and agent interact with hardware
exclusively through Ophyd, not raw EPICS writes.
Federated authentication is handled by Keycloak~\cite{keycloak}, which
issues tokens that gate both API access and the progressive-disclosure
privilege levels described above.
Inter-process communication (IPC) runs over NATS \cite{nats}, a
lightweight publish/subscribe message broker whose role as
the beamline-wide message fabric is detailed in \S\ref{subsec:ipc}.
This stack was not assembled from scratch for Lightfall; every component has
a proven track record, most of them in synchrotron contexts, and our
contribution is the uniform API layer, the IPC abstractions, and the
agent integration that sit above it.

\subsection{Uniform addressability}\label{subsec:addressability}

We use the term \emph{addressable} in a precise sense: a concept class is
addressable when any client (human-facing GUI, script, or agent) can
discover its available operations, invoke them with typed parameters, and
receive structured results or event streams through a single, stable
interface without requiring client-specific adapter code.
In Lightfall, three concept classes meet this definition: panels, devices,
and plans.
Panels expose a discoverable set of methods and properties: selection
state, the controls currently visible in the panel, and event streams that
fire when the panel's state changes.
Devices expose their command surface via Ophyd: every device supports
\texttt{set}, \texttt{read}, and \texttt{subscribe} semantics regardless
of whether the underlying hardware is a stepper motor or a two-dimensional
photon-counting detector.
Plans expose their full parameterization (argument names, types, and
defaults) together with lifecycle controls (pause, resume, abort) via
Bluesky's document model.
All three concept classes expose this surface through a consistent
introspection API (an interface through which any client can enumerate, at
runtime, the operations an object makes available): the same call that the
GUI uses to enumerate a panel's
controls is the call the agent uses to populate its tool list.
The GUI is, in this respect, one client among several; it has no
privileged access to information that is unavailable to any other client
that holds the appropriate authentication token.

This uniformity is the system's central enabling property.
Because panels, devices, and plans all present the same addressable
surface, a single agent that can reach that surface can act on all three
concept classes without any special plumbing per concept type.
There are no bespoke tool schemas per device family and no hand-written
agent adapters per panel type; the agent's tool definitions are generated
at runtime from the same API metadata the GUI consumes.
This generated schema covers the three core concept classes; agent
plugins (\S\ref{subsec:plugins}) can add hand-authored tools alongside
it, but reaching panels, devices, and plans requires no per-type adapter
code.
When a new panel plugin is installed, the agent gains the ability to
interact with it automatically, because the plugin registered its
interface against the shared API rather than against the GUI renderer.
The dual-role pattern---one agent acting as both a user and a developer of
the interface---falls out of this design directly: the agent
uses the interface the same way every other client does, and the API's
self-describing character means that new capabilities propagate to the
agent without a separate documentation step.
Section~\ref{sec:llm-designer} demonstrates this property in the design
role, where the agent composes and modifies panels by invoking skills that
themselves register against the same surface;
Section~\ref{sec:llm-user} demonstrates it in the agent's
control role, where it drives an experiment end-to-end through the uniform
interface.

\subsection{Plugin system}\label{subsec:plugins}

Lightfall's runtime is organized around a small open family of plugin types,
each with a defined registration contract; new types can be added without
core changes, but the ones load-bearing for this paper are the following.
\emph{Panel plugins} add UI surfaces: a plugin declares a panel class,
and the runtime instantiates it, wires its event streams to the message
bus, and registers its addressable interface without any modification to
core code.
\emph{Settings plugins} add preference pages that appear in the
facility-wide configuration dialog, allowing beamline-specific parameters
to be surfaced alongside global ones.
\emph{Engine plugins} add acquisition backends (alternative RunEngine
configurations, for example, or wrappers around hardware that does not
speak EPICS), so that beamlines with non-standard hardware can participate
in the platform without forking the core.
\emph{Plan plugins} add reusable scan procedures that appear in the plan
library and are immediately addressable through the uniform API once
registered; a plan plugin written for one beamline can be reused at
another with no changes to the hosting system.
\emph{Agent plugins} are the most consequential plugin type for the
purposes of this paper. An agent plugin extends the embedded LLM with a
\emph{skill} (a markdown system-prompt contribution, optionally
accompanied by reference documents loaded on demand) and/or with a set of
callable MCP tools registered in-process. Both contributions take effect
at startup or on activation without restarting the application or
modifying the base model.
This shape is intentionally close to recent ``skill''-style packaging
patterns in agent frameworks: the authoring surface is instructions and
reference material rather than a bespoke runtime, and the optional tool
bag is a thin addition rather than the primary deliverable.
Agent plugins are discovered at startup and can be scoped per beamline,
so the agent at one beamline does not automatically acquire capabilities
that are irrelevant or unsafe at another. The growing tool inventory
(panel-design and panel-builder skills, device-catalog tools, plan-launch
tools, plotting and logbook bindings) is added alongside the existing
ones as new MCP tool sets, not as modifications to the agent core.
The distinction between control-mode and design-mode agent plugins maps
directly onto the two LLM roles the paper's thesis claims: control-mode
plugins teach the agent to execute beamline-specific experimental
procedures, while design-mode plugins teach it to extend the interface
itself.

\subsection{Inter-process communication and resource addressing}\label{subsec:ipc}

The uniform addressable surface described above is not confined to the
Lightfall application. Lightfall's runtime participates in a beamline-wide
message bus, implemented on NATS \cite{nats}, over which panels, devices,
and plans publish state changes and accept invocations as pub/sub
subjects.
NATS was chosen over alternatives such as MQTT, ZeroMQ, Kafka, and Redis
for four practical reasons: built-in request/reply semantics, so a client
can invoke an endpoint and receive a direct response without extra
messaging plumbing; single-binary deployment with native TLS; hierarchical
subject
routing with wildcards that maps cleanly onto facility/beamline
namespaces; and mature multi-language client coverage so that
participants written in any language can join the bus without an
adapter layer.
Every addressable entity has a canonical subject address under
a beamline-scoped topic prefix: \texttt{<prefix>.meta.actions} enumerates
the callable actions a Lightfall instance registers on the bus,
\texttt{<prefix>.meta.events} enumerates the event streams it emits, and
per-entity subjects carry typed state changes as they occur. This is the
resource-addressing pattern familiar from IoT messaging: each resource is
a named endpoint on the bus that any authenticated participant can
subscribe to or send to.

Instance discovery is explicit. Any Lightfall instance on the bus responds
to a \texttt{\_lightfall.discover} broadcast with its identity, display
name, and the topic prefix under which its actions live, so a client
joining the bus can enumerate the available Lightfall instances, fetch each
one's action schema, and invoke actions through request/reply patterns
without opening a socket to the Lightfall process itself. The bus is
TLS-protected. Authorization for external clients is granted through an
explicit per-session trust handshake on a dedicated subject
(\texttt{auth.request}): the external client identifies itself by name
and version; the running Lightfall instance prompts the human user to
approve or deny (with a short auto-reject timeout); and on approval the
user's session tokens are returned through the requester's ephemeral
reply-inbox rather than over a public subject. A trusted client then
acts under the granted authorization for the lifetime of the Lightfall
session.

The bus's bidirectional, low-latency request/reply is what lets external
services participate in closed-loop experiments. In the canonical case,
a Lightfall instance publishes Bluesky event documents on its event subjects
as a scan progresses; a live-analysis service subscribes to that stream
and computes a derived signal (an alignment metric, a peak position, a
correlation function from a coherent-scattering measurement), applying
the on-the-fly data-assessment pattern of Ren
et~al.~\cite{ren2017onthefly} to the streaming document model; an
autonomous experiment engine consumes the analysis output, evaluates a
surrogate model, and posts plan-parameter suggestions back to Lightfall's
action subjects via request/reply; Lightfall re-invokes its plan with the
suggested parameters, and the loop closes. None of these steps require
modifying Lightfall's core: each participant addresses Lightfall through the
same uniform surface that the GUI and the embedded agent use, and each
joins or leaves the loop independently of the others. This is the
property that lets analysis, autonomous decision-making, and acquisition
co-evolve on a beamline without re-architecting the control system.

The consequence is that the addressability property that makes the
embedded agent's job tractable is the same property that opens the
system to external processes. A Python script, a service on a
neighboring machine, and a third-party tool on a scientist's laptop
address the same things the GUI does; the architecture treats them as
peers rather than as special cases. Section~\ref{subsec:ecosystem}
reports the external clients that already exercise this surface today,
including the autonomous engine described above and an external Anthropic
CLI agent reaching in through an MCP-to-NATS bridge.

The next two sections examine each agent role in turn.

%% file: content/04-llm-as-designer.tex
\section{LLM-as-designer: runtime extension via skills}\label{sec:llm-designer}

Because panels, devices, and plans share the addressable interface described
in \S\ref{sec:architecture}, an embedded agent (a hosted LLM running
in-process with the application; \S\ref{sec:llm-user} details how its tool
schema is generated) can not only \emph{use} those objects
as an experiment operator but also \emph{compose and modify them} as an
interface developer.
This section demonstrates the latter role, the paper's headline result: an
agent receiving a natural-language
request from beamline staff, invoking a panel-design skill, and producing a
concrete, version-controlled edit to the running interface.
Section~\ref{sec:llm-user} returns to the former role, where the same agent
drives an experiment end-to-end through the same surface.
The mechanism is the same API-first property applied to a different class
of action.
An agent that can query a panel's current state can, given the appropriate
skill, also propose a change to that state's definition.
The uniformity of the surface is what makes both roles coherent without
per-role engineering.

It is worth being precise about what design mode is and is not. Design mode
is more than a developer running a coding agent against the Lightfall source
tree from the outside. The agent runs in-process inside the live Lightfall instance and is
embedded in the beamline ecosystem rather than in a separate development
workspace. It can read the live application's state through the same
introspection API the GUI uses, observe rendered panels and live device
readings as a user would, query the running device catalog and the
Keycloak-gated permission boundaries, and exercise the message bus
(\S\ref{subsec:ipc}) against real or simulated hardware. Modifications
that it makes are loaded into the running application on the fly, so the
agent (and the user it is working with) can verify each change
against the live system within the same session. Platform features
(authentication, the device registry, the logbook, the Tiled catalog)
are themselves first-class tools to the agent, not external services it
has to be taught about. The collaborator at the keyboard is most often a
beamline scientist using their own interface, not a software developer
operating on the codebase from outside.

An AgentPlugin, introduced in \S\ref{sec:architecture}, is a runtime
plugin with two complementary responsibilities, either of which may
appear on its own. It may contribute a \emph{skill}: a markdown
system-prompt snippet plus, optionally, supplementary reference documents
that the agent loads on demand at the point a relevant task arises. And
it may contribute a small bundle of callable \emph{tools}, registered
in-process as MCP tools against the agent's tool schema. A pure-skill
plugin carries instruction and reference material with no new tools (the
panel-design skill described below is one example); a pure-tool plugin
adds device or plan tools without changing the agent's prompt-level
instructions; many
plugins do both, in which case the prompt describes when and how to
reach for the accompanying tools.
This shape is deliberately close to the recent ``skill''-style packaging
patterns in agent frameworks (most directly Claude Skills), and the
authoring surface looks much more like writing instructions and reference
notes than like writing a Python application: the connective tissue (the
plugin class, the registration glue, the on-demand reference loader) is
provided by the runtime.
Both contributions take effect at startup or on activation without
restarting the application or touching the base model.
Agent plugins are discoverable at runtime through the same introspection
mechanism used by panel and plan plugins: the runtime enumerates the
registered plugins, reads their system-prompt contributions and tool
signatures, and assembles the composite system prompt and tool list that
govern the agent's behavior for a given session.
A plugin can be activated or deactivated per session, so a beamline that
shares hardware between multiple experimental programs can present a
different agent vocabulary to each program without maintaining separate
deployments.
Plugins live under version control in the beamline's repository; beamline
staff own that repository and author, review, and merge plugins using the
same git workflow they apply to any other software asset.
The LLM capability is a runtime consequence of that repository's
contents, not a separate system maintained by a specialist.

Panel-design support is split across a small family of cooperating agent
plugins.
Two of them sit at the center: \texttt{panel\_design} is prompt-only and
contributes API-reference material covering \texttt{BasePanel},
\texttt{PanelMetadata}, the panel lifecycle, and the self-registration
pattern that Lightfall uses for user plugins, with the bulk of that material
shipped as on-demand reference documents loaded only when the agent
reaches for them; \texttt{panel\_builder} contributes the MCP tools that
actually write and manage those plugins on disk.
A separate \texttt{device\_tools} plugin is a pure-tool agent plugin
(no system prompt of its own) that provides catalog introspection through
tools such as \texttt{lightfall\_list\_devices}, \texttt{lightfall\_get\_device}, and
\texttt{lightfall\_get\_catalog\_info}, which return structured information
about registered hardware and let the agent ground natural-language
references to concrete signals.
A staff user invokes the family by expressing their intent in the agent
chat interface: ``add a temperature readout to the alignment panel,'' or
``move the beam-stop control to the top of the hazard panel.''
The principal builder tool, \texttt{lightfall\_create\_user\_plugin}, accepts a
complete plugin source file, validates it (syntax check, exec-time
import check, and discovery of at least one concrete
\texttt{PluginType} subclass), writes it to the user-plugin directory
(\texttt{\textasciitilde/lightfall/plugins/}), and loads it into the
running application.
Companion tools handle related operations:
\texttt{lightfall\_create\_temp\_plugin} for ephemeral prototypes that vanish
on restart, \texttt{lightfall\_list\_user\_plugins} for inventory,
\texttt{lightfall\_reload\_plugin} to force a reload after an external edit,
and \texttt{lightfall\_unload\_plugin} to remove a plugin from the registry.
For targeted modifications to an existing plugin (relabeling a
widget, repositioning a control, swapping a device address), the agent
operates on the plugin source file directly with the editor tool from its
generic toolbelt and follows up with \texttt{lightfall\_reload\_plugin} when a
hot reload is required.
The review surface is the chat pane plus the per-tool permission hooks:
the agent's proposed code is visible to the staff user before any tool
invocation proceeds, and a hook on each tool gives the user a final
chance to refuse.
The validation step in \texttt{lightfall\_create\_user\_plugin} (syntax,
import-time, and \texttt{PluginType}-discovery checks) gates whether the
file ever lands.
Once it does, the create and modify tools record the change as a git
commit in the user-plugin repository, so the diff between successive
commits is the canonical structural record of what the agent did,
inspectable through the same tooling beamline staff already use to
review hand-written changes.

\begin{figure}[tbp]
  \centering
  \includegraphics[width=0.95\linewidth]{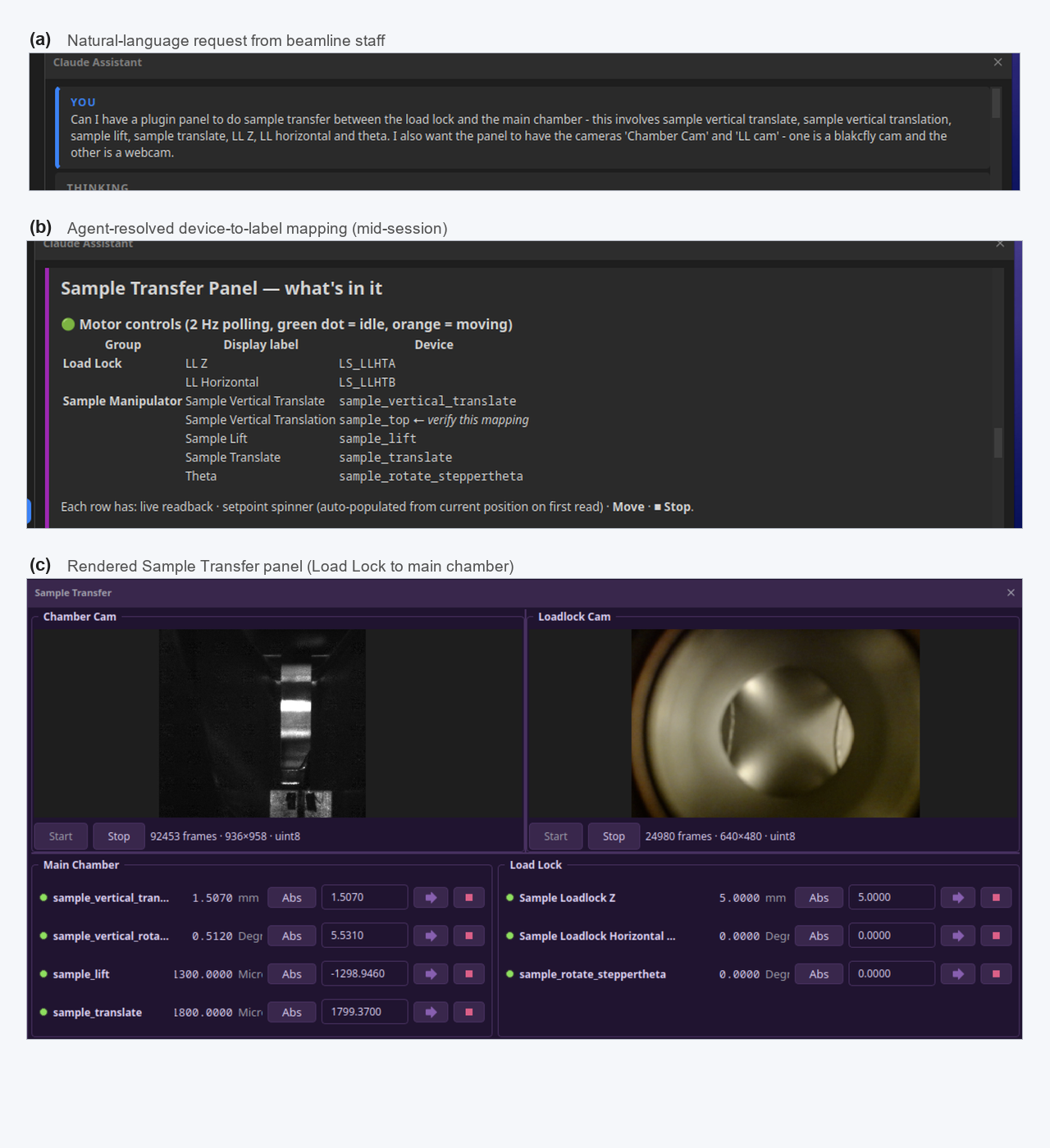}
  \caption{LLM-as-designer at COSMIC. (a) A beamline scientist's
  natural-language request for a new Sample Transfer panel. (b) The
  agent's proposed mapping from natural-language axis names to registered
  device identifiers. Some of the assumed mappings are incorrect: one
  ambiguous axis is self-flagged for user verification
  (``\texttt{sample\_top} $\leftarrow$ verify this mapping''), while
  another (``LL Z'' to \texttt{LS\_LLHTA}) is misassigned without being
  flagged.
  (c) The rendered panel hot-reloaded into the running Lightfall instance,
  with two live camera views and two motor groups that the agent organized
  into sample-manipulation controls on the left and load-lock controls on
  the right, shown after a short
  follow-up exchange in which the scientist's feedback corrected the
  flagged and unflagged device mappings (sample vertical rotation in place
  of the agent's tentative guess) and the agent re-emitted the plugin.}
  \label{fig:design-mode}
\end{figure}

Panel-design skills cover panel creation in addition to modification of
existing panels.
To make this concrete, we reproduce a session recorded at COSMIC on
2026-05-04 in which a beamline scientist requests a new panel from
scratch (Figure~\ref{fig:design-mode}).
The scientist types: ``Can I have a plugin panel to do sample
transfer between the load lock and the main chamber --- this involves
sample vertical translate, sample vertical translation, sample lift,
sample translate, LL Z, LL horizontal and theta. I also want the panel to
have the cameras `Chamber Cam' and `LL Cam' --- one is a Blackfly cam and
the other is a webcam.''
The agent, with the \texttt{panel\_design} and \texttt{panel\_builder}
skills active, calls \texttt{lightfall\_list\_devices} to enumerate the
registered catalog and inspects the camera registry, then proposes a
mapping from the user's natural-language axis names to concrete device
identifiers (for example, ``Sample Lift'' to \texttt{sample\_lift}).
Not every assumed mapping is correct: ``LL Z'' is initially assigned to
\texttt{LS\_LLHTA}, an incorrect identification that the scientist later
corrects through feedback, and one axis name, ``Sample Vertical
Translation'', has no unambiguous match
in the registry; the agent flags it as a tentative mapping rather than
guessing silently.
The agent then assembles a complete plugin source file with two motor
groups (Load Lock and Sample Manipulator), per-axis live readback at
2~Hz, a top-of-panel emergency-stop control, and two camera widgets
bound to the Blackfly GigE address and the local webcam.
It emits the file through \texttt{lightfall\_create\_user\_plugin}, which
validates the code, writes it to the user-plugin directory, and loads it.
The panel registers and renders on the next hot reload
(Figure~\ref{fig:design-mode}).
A short follow-up exchange refines the result: the scientist corrects both
the flagged mapping (``the correct device is sample vertical rotation'')
and the unflagged \texttt{LS\_LLHTA} misassignment,
updates the Blackfly IP address after a network reconfiguration, and asks
the agent to install OpenCV into the Lightfall virtual environment when the
LL Cam first reports \texttt{cv2} unavailable.
Each follow-up is a small, single-scope edit applied and verified before
continuing.

We evaluated this workflow through informal sessions with beamline scientists at COSMIC during the testing period described in
\S\ref{sec:deployment}, gathering on the order of 10--20 panel-design
requests across the scope categories described below.
Sessions were loosely structured: some requests arose spontaneously during
normal operation, others were suggested as directed tasks; in all cases a
beamline scientist formulated the request in their own words, and a Lightfall developer was present to observe the agent's response and
intervene if the interaction diverged from what the scientist had intended.
The observations reported below are qualitative and should be read as such.

In terms of scope, requests fell into four categories: adding or
repositioning readout widgets (temperature sensors, pressure gauges,
beam-position monitor channels), reorganizing button groupings within a
panel, adding a live-plot pane bound to an existing Tiled data stream,
and adjusting parameter defaults on plan controls.
The interesting observation across these sessions is not a
correctness percentage in the usual sense.
The pattern that emerged repeatedly is that the beamline scientist becomes
the author of the change.
Initial requests were almost always under-specified in ways that the
scientist did not notice until the rendered panel surfaced the gap:
which controls should sit together, whether a readout should be live or
on-demand, what axis names map to which devices, where the panel should
appear relative to existing surfaces. Each of these is a requirements
question, not a model failure, and the natural response is iteration. In
this workflow that iteration runs inside the same chat session with the
same agent against the same live application: the scientist refines the
specification, the agent re-emits, and the panel updates in place.
Crucially, the scientist does this themselves, without scheduling time
with a core developer and without leaving the running interface.
The development cycle closes inside the session: the scientist
authors the change in natural language, the agent emits and applies
it, the panel renders inside the live application, and the commit
lands in the beamline's plugin repository as a side effect,
without acquiring a developer review queue, a redeploy step, or any
hand-off outside that repository.
This is the property that makes the workflow scale.
The marginal cost of a per-scientist iteration is fixed in the
scientist's own time rather than in developer hours the facility must
allocate; developer effort scales with the platform itself
(plugin APIs, agent skills, tool surfaces) instead of with the
number of scientist-driven iterations that depend on it.
Propagation of a change to broader facility use still passes through
normal review, as described below, but the per-scientist iteration
loop no longer waits on it.
What the panels can express is, in practice, bounded mainly by how
precisely the requirements are stated rather than by the agent's reach:
when the request is unambiguous, the first emission is typically
structurally complete and renders cleanly; when the request leaves a
question open, the panel reflects an answer, and the scientist refines
from there.
The same dynamic governs panel creation from scratch
(Figure~\ref{fig:design-mode}): the first emission was structurally
complete and rendered, and the device mappings and network addresses
that still needed attention were precisely the requirements details a
core developer would have had to ask about anyway, now asked and
answered in a single session.
A handful of cases did require multiple turns: multi-panel modifications
in which changes to one panel's layout implied related changes to
another, and modifications that needed a new device binding because the
target signal had not previously been registered in the panel
configuration.
Even these resolved within the session rather than across a separate
development cycle.

The git-backed commit model is what makes this workflow acceptable under
normal facility software governance.
Every agent-produced panel change is recorded automatically as a commit
in the beamline's skills and configuration repository: the create and
modify tools wrap each successful write in a \texttt{git add} and
\texttt{git commit} step, so the auditable history is built up as a
side effect of normal use rather than a separate staff workflow.
That means every change is visible in pull-request review before it is
merged to the branch from which the deployment server reads; normal
code-review discipline applies, and no agent-generated change can reach
other users without passing through that gate.
Every change is revertible with standard \texttt{git revert}, without any
Lightfall-specific tooling.
Facility staff retain veto power at the review step: AI assistance
accelerates the authoring of a change but does not alter the path that
change takes through the software-release process.
The cumulative commit history is itself a record of how the interface
evolved under real use: who requested what, when, and why. It parallels
the Tiled data provenance described in \S\ref{sec:supporting}.
This is the sustainable middle ground introduced in
\S\ref{sec:introduction}. AI assistance amplifies staff throughput without
removing human review, and it does so using infrastructure (version
control, code review, git history) that beamline staff already understand
and trust.

%% file: content/03-llm-as-user.tex
\section{LLM-as-user: control mode}\label{sec:llm-user}

Because panels, devices, and plans share the uniform API described in
\S\ref{sec:architecture}, the embedded agent can be wired to that surface
without any per-beamline plumbing.
The agent is a hosted language model (Claude Sonnet 4.6 in the deployment
reported here) running in-process with the Lightfall application.
Its tool schema is auto-generated at startup from the same introspection
API the GUI uses: every registered panel, every Ophyd device, and every
plan plugin contributes its callable methods, typed parameters, and
permissible argument ranges to the agent's tool list without a separate
documentation or hand-maintenance step.
When a new plugin is installed, the agent gains access to its interface
automatically, for the same reason the GUI does: both are clients of the
shared API surface.
This approach follows the tool-use pattern established in the broader
agent literature \cite{yao2023react,schick2023toolformer,anthropic-tool-use,xi2023agents},
with the distinction that the core tool catalog is not curated by hand
but derived structurally from a self-describing runtime; hand-authored
tools enter only as agent-plugin contributions (\S\ref{subsec:plugins}).

\begin{figure}[tbp]
  \centering
  \begin{subfigure}{0.95\linewidth}
    \centering
    \includegraphics[width=\linewidth]{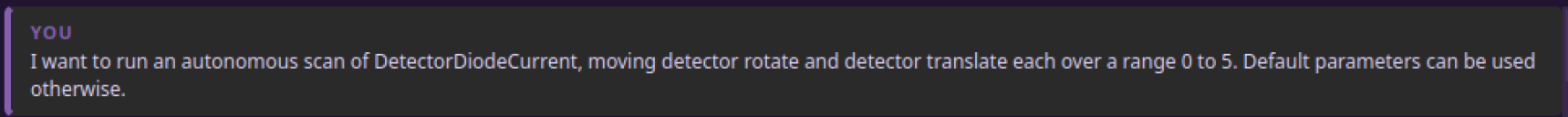}
    \caption{The scientist's natural-language request for an autonomous
    scan.}
    \label{fig:control-mode-a}
  \end{subfigure}\par\medskip
  \begin{subfigure}{0.95\linewidth}
    \centering
    \includegraphics[width=\linewidth]{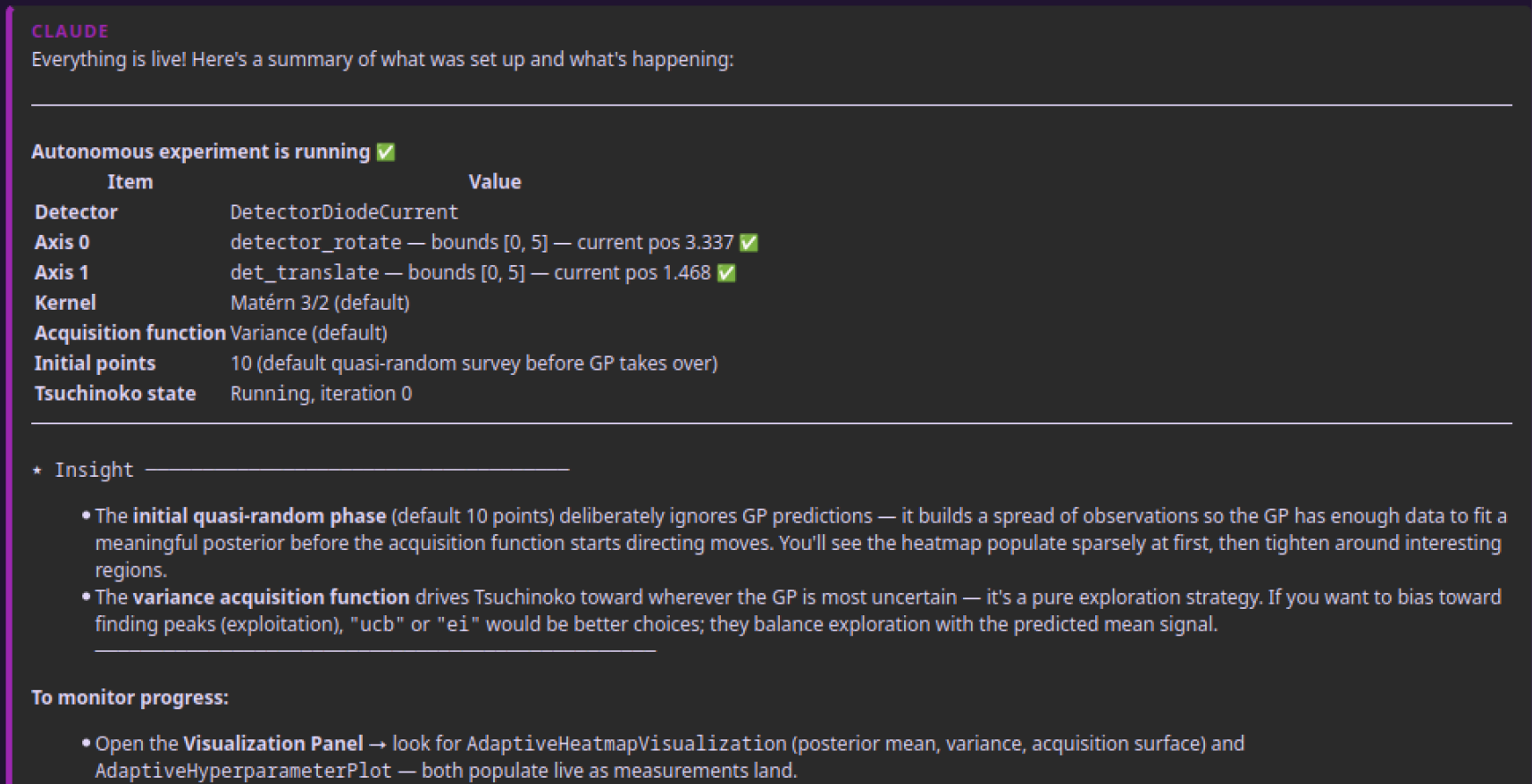}
    \caption{The agent's structured launch summary: the detector and
    axes resolved from the request, the defaults adopted for kernel,
    acquisition function, and initial-point count, and pointers to the
    Tsuchinoko visualization panels for live monitoring.}
    \label{fig:control-mode-b}
  \end{subfigure}\par\medskip
  \begin{subfigure}{0.95\linewidth}
    \centering
    \includegraphics[width=\linewidth]{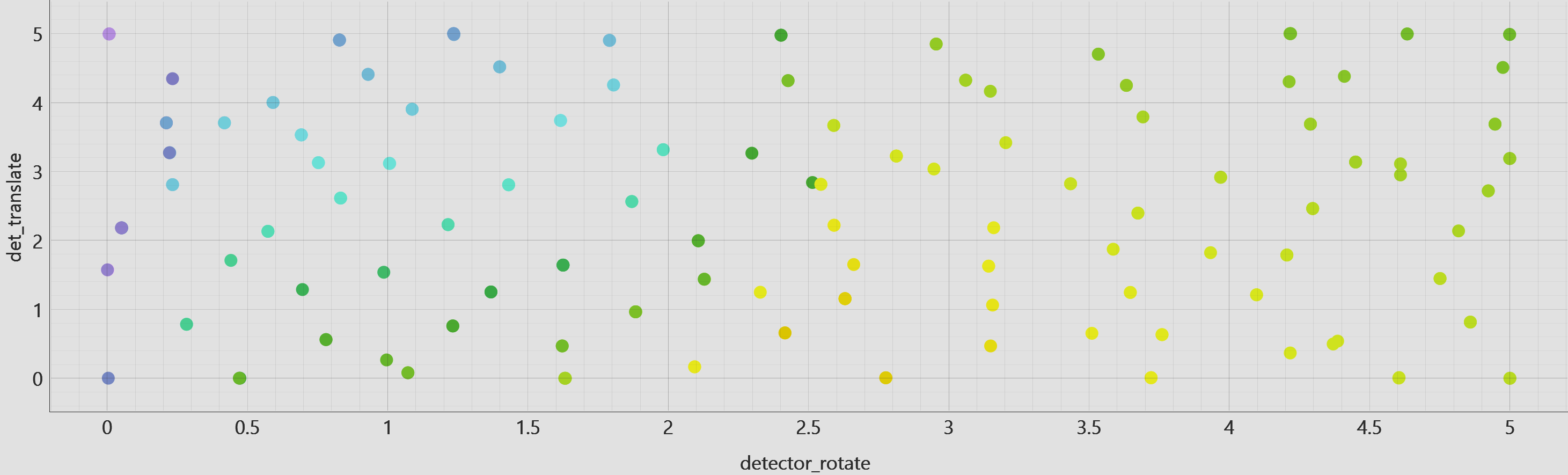}
    \caption{Tsuchinoko visualization of the autonomous scan in
    progress: the sampled positions in the
    (\texttt{detector\_rotate}, \texttt{det\_translate}) plane, with
    color encoding the measured \texttt{DetectorDiodeCurrent} value
    at each point. The variance acquisition function (the run's
    default) drives the sampler to spread points across the parameter
    space rather than cluster around an extremum, which is the
    intended behavior during pure exploration.}
    \label{fig:control-mode-c}
  \end{subfigure}
  \caption{Embedded agent in control mode: an autonomous scan
  dispatched from a single natural-language request. The agent
  resolves the scientist's intent to a short sequence of calls against
  Tsuchinoko (\S\ref{subsec:ecosystem}) and the registered
  \texttt{adaptive\_experiment} plan; the Tiled run UID and full
  tool-call record are persisted to the logbook for later
  reconstruction.}
  \label{fig:control-mode}
\end{figure}

To make this concrete, we walk through a session recorded on 2026-05-19
at the COSMIC-Scattering beamline, with the agent driving real beamline
hardware. A scientist types: \emph{``I want to run an autonomous scan
of \texttt{DetectorDiodeCurrent}, moving \texttt{detector\_rotate} and
\texttt{det\_translate} each over a range 0 to 5. Default parameters
can be used otherwise''} (Figure~\ref{fig:control-mode}a).
The agent discovers the running Tsuchinoko instance (the headless
adaptive-experiment service of \S\ref{subsec:ecosystem}) via
\texttt{tsuchinoko\_discover}, sends it the experiment design with
\texttt{tsuchinoko\_configure}, and dispatches the registered
\texttt{adaptive\_experiment} plan against it: detector
\texttt{DetectorDiodeCurrent}; two axes \texttt{detector\_rotate} and
\texttt{det\_translate}, each bounded to $[0,5]$; defaults for the
remainder (Matérn $3/2$ kernel, variance acquisition, ten quasi-random
initial points before the Gaussian-process posterior takes over).
It returns a structured launch summary
(Figure~\ref{fig:control-mode}b) and points at the Tsuchinoko
visualization, which populates live as measurements land
(Figure~\ref{fig:control-mode}c).
The Tiled run UID and the full tool-call record are written to the
beamline logbook as a single timestamped session entry, so the scan is
reconstructible from the record alone.

Access control and audit are enforced at the API layer.
Tool availability is gated by the Keycloak token held by the current
agent session: a staff-authenticated session exposes device set calls and
plan abort capabilities that are absent from a user-level token.
High-stakes actions require an explicit in-band confirmation (a
yes/no prompt presented to the operator in the same chat session) before the
agent is permitted to issue the tool call. These include motor moves
exceeding a configurable displacement threshold and destructive plan
transitions such as abort-and-reset.
Every agent invocation and every tool call is written to the logbook
with its full parameter record and the authenticated principal that
triggered it, so an experiment session is fully reconstructible from the
record without relying on operator memory or screen recordings.
The agent and its tool layer are not part of the equipment- or
personnel-safety envelope. Protection against hardware damage and
operator injury is the responsibility of lower layers: the device
controllers (motion limits, soft and hard stops, IOC-level command
validation) and the independent hardware safety systems governing
radiation and access enclosures. Both refuse unsafe commands
regardless of origin (LLM, human operator, or any other API client).
The controls described here are engineering controls that reduce
inadvertent operator-visible mistakes and provide accountability after
the fact; they are not a primary safety mechanism.

The same control-mode pattern is also realized by external agents
running outside Lightfall. A scientist driving Claude Code, an
IDE-integrated agent, or a custom MCP client from their own machine
reaches a running Lightfall instance through \emph{lightfall-mcp-bridge}, an
MCP server that subscribes to Lightfall's addressable surface over NATS
(\S\ref{subsec:ipc}) and re-exposes each registered action as an MCP
tool. Because the bridge presents the same uniform API, an external
agent sees the same tool schema, operates under the same Keycloak-gated
permissions, and produces the same auditable invocation trail in the
logbook as the in-process agent. The two realizations differ only in
where the agent runs, not in what it can do.
Section~\ref{subsec:ecosystem} reports the realized external client.

Control mode handles routine beamline operations well.
A query such as \emph{``what is the current sample temperature?''}
resolves to a single \texttt{lightfall\_read\_device} call.
Scan setup, parameter sweeps, and acquisition status summaries are
similarly well-served: the agent's tool schema covers the full uniform
API surface. What the agent can actually issue unattended is a subset,
shaped by the safety gates described above.
Highly novel or open-ended experiment planning, and fault diagnosis
under ambiguous conditions, remain beyond reliable agent performance
today; these are future-work items discussed in \S\ref{sec:discussion}.
The benefit of the API-first design is that improving the agent's
reasoning, by upgrading the underlying LLM or extending skills, does not
require any changes to the control system itself.

%% file: content/05-supporting.tex
\section{Supporting capabilities}\label{sec:supporting}

The architectural property that \S\ref{sec:llm-designer} and~\S\ref{sec:llm-user}
exploited, a uniform addressable API surface across panels, devices, and
plans, carries consequences beyond the two LLM roles.
Two supporting capabilities emerge from the same design without additional
machinery: automatic FAIR data cataloging (which also extends naturally to
remote operation through the same authenticated control plane) and a coherent
facility-wide user experience (UX).
A growing external client ecosystem (\S\ref{subsec:ecosystem}) exercises
the same surface from outside the main application.
Each is briefly described below, with the connection to the API-first claim
made explicit.

\subsection{FAIR data via Tiled}\label{subsec:fair}

Because every scan plan is an addressable API object (\S\ref{sec:architecture}),
every run executed through Lightfall is automatically submitted to Tiled
\cite{tiled}, the data-catalog service introduced in
\S\ref{subsec:core-stack}, with its full parameter set, data arrays, and
provenance metadata.
The FAIR principles \cite{wilkinson2016fair} (findable, accessible,
interoperable, and reusable) are supported by construction through Tiled
and the Bluesky document model, though the degree to which each principle
is met depends on beamline practice.
Findability and accessibility are direct: runs are findable through Tiled's
query API by plan type, device name, time range, or any parameter recorded
in the document model, and they are accessible via standard HTTP, making
post-acquisition browsing and programmatic re-analysis available to
collaborators with no Lightfall installation required.
Interoperability and reusability follow to the extent that beamlines
populate the document model's metadata fields with controlled vocabularies:
the Bluesky document model \cite{allan2019bluesky} provides a shared
serialization across facilities, and the plan parameters that produced a
run are stored alongside its data, so any run can be re-invoked without
reconstructing intent from notes.
Authentication of data access is unified with the control plane: Lightfall's
custom authenticator for Tiled validates Keycloak JWTs directly, so a
user's session authorizes both control operations and data retrieval
through the same token. Read access to a run is therefore governed by
the same identity that authorizes motor actuation, and access rights
follow the user across Lightfall's GUI, an independent notebook, or a
third-party client without a separate login step.
Broader FAIR guarantees (formal vocabularies, cross-repository linkage,
rich provenance chains) remain subject to community conventions beyond
Lightfall's core; a comparable treatment in the materials-science domain is
discussed in \cite{draxl2019nomad}.
The connection to \S\ref{sec:architecture} is direct: the same addressability
that let the LLM-as-user agent query and invoke plans also makes every
plan's parameters part of the persistent record.
Data provenance is, in this view, the audit trail of addressable
API calls.

The same single-token model extends to remote operation: a Keycloak-authenticated
session retains its scope, safety gates, and audit trail regardless of where the
client connects from, because those properties are enforced at the API layer rather
than at the workstation. Broader remote operation for off-site collaborators awaits a
facility-supported edge proxy and is left to future work.

\subsection{Unified UX}\label{subsec:ux}

Lightfall's plugin system (\S\ref{sec:architecture}) inverts the usual failure mode
for shared facility software.
At most facilities, the shared codebase is what gets copied: each beamline forks
a GUI template, customizes locally, and eventually diverges to the point that
upstream improvements cannot be merged.
Because Lightfall panels register against the shared API rather than against the
renderer, facility-wide consistency is the default state and per-beamline
customization is the override.
A theming layer applies facility-wide look-and-feel; beamlines supply an
override file to adjust colors, logos, and default layouts without touching
shared code.
Progressive disclosure, with routine controls visible to all users and
privileged or experimental panels gated behind Keycloak roles, is enforced
at the API layer, so the disclosure policy is uniform regardless of which
beamline or which client is rendering the panel.
Per-user preferences (units, layout state, preferred plan parameters) are stored
against the authenticated user identity and persist across sessions and machines.
These properties hold across all deployed beamlines because they derive from
the shared API contract, not from coordination between forks.

\subsection{External client ecosystem}\label{subsec:ecosystem}

Because the message bus of \S\ref{subsec:ipc} is open to any authenticated
participant, Lightfall's API is already exercised by clients outside the main
application. Three examples are worth highlighting.
\emph{Tsuchinoko}, a headless adaptive-experiment service, subscribes to
device and plan events and publishes plan-parameter suggestions back to
the bus, letting a Gaussian-process surrogate~\cite{noack2021gaussian,risser2025nonstationary}
co-pilot an experiment without modifying the Lightfall runtime.
The \emph{als-tiled-exporter} subscribes to acquisition documents on the
bus and writes them into the Tiled catalog, stamping each document with
the authorization metadata that downstream access control depends on
(\S\ref{subsec:fair}); the data archive therefore stays current with live
acquisition, and Tiled becomes the canonical FAIR endpoint rather than a
post-hoc upload target.
\emph{Claude Code}, the Anthropic CLI agent, reaches into Lightfall through
\emph{lightfall-mcp-bridge}, an MCP server that subscribes to a running
Lightfall instance's NATS subjects (\S\ref{subsec:ipc}) and re-exposes each
registered action as an MCP tool inside the scientist's terminal
session. The bridge runs on the scientist's machine and is authorized through
the per-session trust handshake described in \S\ref{subsec:ipc}, so the
external agent operates under the user's session tokens with the same
scope, the same per-tool permission gates, and the same logbook audit
trail as the in-process agent of \S\ref{sec:llm-user}; it is a second
realization of the same pattern, not a privileged side channel. The realization is useful in practice
when the scientist is already working in a terminal-driven environment
(a Jupyter session bound to a remote kernel, an IDE-integrated
agent, or a scripted batch workflow that wraps several beamline
operations into one composite task) and prefers to drive Lightfall from
there rather than through the embedded chat pane. The same uniform API
supports both agent hosts on equal footing.
Together these tools are evidence that the addressability claim is not a
design aspiration: the API is doing real work for multiple independent
clients today.

%% file: content/06-deployment.tex
\section{Deployment at COSMIC-Scattering}\label{sec:deployment}

COSMIC-Scattering, one of the two endstations of the COSMIC beamline at
the Advanced Light Source, Lawrence Berkeley National Laboratory, supports
X-ray photon correlation spectroscopy (XPCS) and related coherent scattering
techniques; recent work at the endstation~\cite{tumbleson2025nematic,singh2024ergodicity,singh2023temporal}
typifies the magnetic-dynamics studies it enables.
The sibling endstation, COSMIC-Imaging, hosts ptychographic imaging and is
not the subject of this report.
A typical XPCS experiment acquires extended time series of detector frames
while the sample dynamics evolve; subsequent intensity-correlation analysis
yields the temporal correlation functions that characterize those dynamics.
Operations span sample alignment, scan-parameter selection (step sizes, exposure
times, detector configuration), and continuous monitoring of live data quality
during the scan.
The diversity of decisions made in a single session, from motor
positioning to on-the-fly parameter adjustment, makes COSMIC a representative
context in which to evaluate Lightfall's integrated capabilities.

Lightfall has been deployed at COSMIC-Scattering in a testing configuration since
early 2026.
The components running at the beamline include the Lightfall core server, Bluesky
acquisition with Ophyd-abstracted devices~\cite{allan2019bluesky,ophyd}, a Tiled
catalog for acquired data~\cite{tiled}, the embedded agent with panel-design
skills, and logbook integration.
Figure~\ref{fig:cosmic} shows the full application as deployed:
device panels, the embedded agent, and the logbook threaded through a
single audit trail.
This is an honest status report: the system is in active testing rather than
full production, and the operational configuration continues to be refined as
real experiments run against it.
The deployment is intended to exercise all three of the capabilities described
in \S\ref{sec:supporting} under realistic facility conditions.

A composite session, assembled from observations during the testing period,
illustrates how the pieces work together.
A beamline scientist arrives for a scheduled experiment, opens the Lightfall
interface, and uses the embedded agent in control mode (\S\ref{sec:llm-user})
to align the sample: the agent sequences motor moves and reads back diagnostic
signals via natural-language requests, issuing the corresponding Bluesky
plans through the structured tool interface.
Once alignment is confirmed, the scientist asks the agent to launch an
XPCS measurement with specified frame rate and total acquisition time; the
agent constructs and submits the plan, and acquisition begins.
Partway through the session the scientist notices that a relevant device
readout is absent from the main panel.
Rather than interrupting the scan to edit configuration files by hand, they
ask the embedded agent to add the readout; the agent invokes the panel-design
skill (\S\ref{sec:llm-designer}), emits the required plugin code, and the
modification is committed to the beamline's panel repository and hot-reloaded
into the running interface. The scan continues uninterrupted throughout.
Post-acquisition, all frames land automatically in the Tiled catalog (\S\ref{subsec:fair})
with full provenance; the logbook (\S\ref{sec:llm-user}) records both the
completed scan and the panel modification with their respective timestamps and
authenticated attribution.

The session described above is representative of what works well in practice.
Routine panel modifications and scan orchestration requests resolve on the first
or second attempt without user intervention beyond the natural-language prompt.
What required iteration was the surrounding infrastructure: initial system setup,
an early round of skill-tuning to match COSMIC-specific conventions (device
naming, preferred plan defaults), and some manual refinement of interface
details that
the agent could not yet handle autonomously.
The qualitative picture is that the architecture imposes no fundamental friction
between the three capability layers (control, design, and data), and that the
effort of adopting the platform is concentrated in initial setup and
configuration rather than
in ongoing operation.

%% file: content/07-discussion.tex
\section{Discussion and future work}\label{sec:discussion}

The current implementation of Lightfall's skill system is text-driven: beamline
staff invoke skills through the chat pane, expressing requests in natural
language, and the agent edits the live application in response.
An open direction is skill authoring itself.
Skills today are authored by developers and merged through the same review
path as any other code change.
A self-hosted authoring workflow, allowing beamline staff to compose new
skills (prompt material, reference notes, and small tool bundles)
through the Lightfall interface without a separate code deployment, would
close the remaining gap between what operators can customize and what
they currently need developer time to achieve. Because agent plugins are
mostly instructions and reference material rather than runtime business
logic (\S\ref{sec:llm-designer}), the authoring surface is closer to a
documentation editor with optional tool wiring than to a code IDE, which
makes a scientist-facing authoring path more plausible than it would be
for a Python-package-shaped plugin.

The autonomous-scan example in \S\ref{sec:llm-user} dispatches a
Gaussian-process-driven adaptive scan, but that is one of several
strategies the agent can choose between through the registered plan
catalog. A small taxonomy is worth stating explicitly: (i) a forward
scan over a parameter range followed by an analytic peak fit and a
move-to-optimum --- the simplest, and the right answer when the
response shape is known; (ii) Bluesky's \texttt{adaptive\_scan}, which
refines step size from the observed slope and is useful for tracing a
feature whose width is not known in advance; (iii) Bluesky's
\texttt{tune\_centroid}, which sweeps past the peak and rescans with
progressively narrower brackets to center a unimodal response; and
(iv) the gpCAM/Tsuchinoko GP scan demonstrated in
Figure~\ref{fig:control-mode}, which builds a posterior over the
response surface and lets a configurable acquisition function direct
each next measurement --- by far the most tunable of the four and the
one that scales naturally to multi-axis surveys.
What the agent surface contributes is that picking between these is
itself a natural-language step: the scientist describes what they are
trying to do, and the agent dispatches the strategy that fits, with
the same authentication, confirmation, and audit guarantees in each
case.

As emphasized in \S\ref{sec:llm-user}, equipment and personnel safety are
the responsibility of the device controllers and the independent hardware
safety systems, not of the agent's tool layer; the discussion here
concerns the operator-visible controls inside that envelope.
The current implementation takes conservative defaults: any tool call
with observable physical consequence (large motor moves, destructive plan
transitions) requires explicit operator confirmation before the agent is
permitted to issue it (\S\ref{sec:llm-user}).
What is not yet in place is a principled, adversarially evaluated policy
that characterizes when the agent can be trusted to act unattended and
under what conditions it should escalate.

A related pattern shows up on the design side, where the safety
surface is a layered combination of catalog grounding, editor-side
guard hooks, and user vigilance rather than a single policy.
A concrete instance from a recent design-mode session at COSMIC
illustrates this combination.
The agent proposed a device-to-label mapping for the Sample Transfer
panel (Figure~\ref{fig:design-mode}) that included a tentative entry
it had self-flagged (``\texttt{sample\_top} $\leftarrow$ verify this
mapping''); the user corrected it on the next turn.
Later in the same session, when the agent attempted a small label
rename, a pre-tool hook blocked the editor call; subsequent fallbacks
through \texttt{lightfall\_create\_user\_plugin} (with overwrite enabled)
and \texttt{lightfall\_ipython\_execute} were also blocked, and the agent
paused for feedback from the user on how to proceed.
Neither outcome is a model failure in the strong sense (the first
is a self-flagged guess, the second a protective denial), but both
illustrate that current design-mode behavior is jointly governed by
the catalog, the hooks, and the user, and that none of the three is
sufficient on its own.
Ongoing work includes red-team-style evaluations of agent behavior around
operationally consequential actions (large motor moves and destructive
plan transitions) and the development of formal confirmation policies
tested under adversarial prompts rather than only under cooperative use.
Formalizing these policies is future work, and we flag it here as an open gap
in the current system's maturity.

Concurrent work occupies two distinct positions.
LLM-driven prototypes such as VISION~\cite{mathur2025vision} and
networked exocortex frameworks~\cite{tsai2025exocortex} share Lightfall's
premise that the LLM is a first-class client of beamline software;
Lightfall's API-first substrate is positioned to interoperate with both,
the in-process agent of \S\ref{sec:llm-user} as one realization of
the LLM-as-user pattern that VISION also pursues, and the open NATS
bus with its trust-handshake authorization (\S\ref{subsec:ipc}) as a
natural attachment point for networked agents.
Build-time component libraries such as Finch~\cite{deleon2025finch}
share the maintainability-through-reuse intuition but keep the
customization loop firmly on the developer side: each beamline
assembles its own application from imported atoms, user-driven
tweaks remain developer tasks, and per-beamline application repositories
accumulate as a deliberate consequence of the design.
Lightfall instead places the customization loop inside the running
application and under the scientist's natural-language control,
treating maintainability as a property of the API surface rather
than of the assembled component set.

%% file: content/08-conclusions.tex
\section{Conclusions}\label{sec:conclusion}

We have presented Lightfall, an API-first control platform for synchrotron beamlines.
A single well-specified API surface makes the system addressable by an LLM
in two complementary roles: as a user of the interface and as a developer of
it. This yields a sustainable alternative to both bespoke per-beamline GUIs
and one-size-fits-all facility software.
The key result is that a single well-specified API surface serves all three principals
simultaneously: the GUI, the embedded control-mode agent, and the design-mode agent
that modifies the interface at runtime through skill invocations.

Relative to comparable beamline-software efforts (bespoke
per-beamline GUIs, facility-wide acquisition shells such as
GDA~\cite{opengda} and CSS-Phoebus~\cite{css-phoebus}, technique-specific
applications such as Blu-Ice~\cite{mcphillips2002bluice} and
MxCuBE~\cite{gabadinho2010mxcube}, and recent component libraries such
as Finch~\cite{deleon2025finch}), Lightfall's contribution is distinct
along five axes that together define its design center.
\begin{enumerate}
\item \emph{Layer of operation.} Lightfall is a running application with
an embedded agent, not a library of building blocks. Customization
happens at runtime against the live application rather than at build
time against a developer's checkout.
\item \emph{Locus of customization.} The customization loop closes with
the scientist running the experiment: natural-language requests
authored, applied, and committed to the beamline's plugin repository
in the same session, without routing through a developer. This is the
property that decides scaling: the marginal cost of a
scientist-driven change is fixed in the scientist's own time rather
than growing linearly in developer effort the facility must supply.
\item \emph{Role of the LLM.} The LLM is a first-class client of the
same uniform API the GUI is a client of, with auto-generated tool
schema, in-band confirmation gates, and logbook-attested audit,
not an external assistant grafted onto a control system that was
designed without it.
\item \emph{Divergence model.} Per-beamline differences are expressed
as plugin configurations and skill bundles loaded by a single
application, rather than as per-beamline application repositories
that must each be maintained, updated, and ported forward.
\item \emph{Scope of the framework.} Authentication
(Keycloak-gated tools), audit (logbook-attested invocations),
autonomous-experiment integration (Tsuchinoko via NATS), FAIR data
access (Tiled), and external-agent reach (the lightfall-mcp-bridge) are
part of the framework, not the beamline application's burden to
re-engineer.
\end{enumerate}

Testing at the COSMIC-Scattering beamline demonstrates the pattern end-to-end.
Lightfall's source is available under a BSD 3-Clause license at
\url{https://github.com/als-controls/lightfall}; we welcome contributions and
collaboration from the wider beamline software community.

%% file: content/09-acknowledgments.tex
\section*{Acknowledgments}

This research used resources of the Advanced Light Source, a U.S.\ Department
of Energy Office of Science User Facility under contract no.\
DE-AC02-05CH11231. This research was also supported by the Center for Advanced Mathematics for Energy Research Applications (CAMERA), which is jointly funded by the Advanced Scientific Computing Research (ASCR) and Basic Energy Sciences (BES) within the Department of Energy's Office of Science, under Contract No.\ DE-AC02-05CH11231.